\documentclass[11pt,preprint]{aastex}
\usepackage{psfig}
\begin{document}

\title{A Ray-Tracing Algorithm for Spinning Compact Object Spacetimes
  with Arbitrary Quadrupole Moments. I. Quasi-Kerr Black Holes}

\author{Dimitrios Psaltis and Tim Johannsen}
\affil{Astronomy and Physics Departments,
University of Arizona,
933 N.\ Cherry Ave.,
Tucson, AZ 85721, USA}

\begin{abstract}
We describe a new numerical algorithm for ray tracing in the external
spacetimes of spinning compact objects characterized by arbitrary
quadrupole moments. Such spacetimes describe non-Kerr vacuum solutions
that can be used to test the no-hair theorem in conjunction with
observations of accreting black holes. They are also appropriate for
neutron stars with spin frequencies in the $\simeq 300-600$~Hz range,
which are typical of the bursting sources in low-mass X-ray binaries.
We use our algorithm to show that allowing for the quadrupole moment
of the spacetime to take arbitrary values leads to observable effects
in the profiles of relativistic broadened fluorescent iron lines from
geometrically thin accretion disks.
\end{abstract}



\section{Introduction}

The observational appearance of accreting black holes and of neutron
stars is strongly influenced by gravitational lensing in their
exterior spacetimes. In the case of accreting black holes, different
aspects of this ray-tracing problem have been addressed by several
research groups to date, as applied to the spectra, timing, and images
of the accretion flows (e.g., Bardeen 1973; Cunningam 1975; Laor 1991;
Rauch \& Blandford 1994; Speith, Riffert, \& Ruder 1995; Dovciak,
Karas, \& Yaqoob 2004; Beckwith \& Done 2004; Broderick 2006; Dexter
\& Agol 2009; Dolence et al.\ 2009).  In the case of neutron stars,
the effect of light bending on the spectra and lightcurves generated
by their surface emission has been explored for non-spinning (e.g.,
Pechenick, Ftaclas, \& Cohen 1983), slowly spinning (Miller \& Lamb
1998; Braje, Romani, \& Rauch 2000; Muno, \"Ozel, \& Chakrabarty 2002;
Poutanen \& Beloborodov 2006), and rapidly spinning neutron stars
(Cadeau et al.\ 2007; Morsink et al.\ 2007).

The external spacetimes of black holes and of slowly spinning neutron
stars are described by the Kerr solution.  This metric is of
Petrov-type D and, therefore, the Carter constant is an integral of
motion along the trajectories of photons (Carter 1968). The presence
of this integral of motion, in addition to the conservation of energy
and angular momentum, allows one to use first-order differential
equations to solve for the photon trajectories. This improves the
accuracy of the calculation and increases its speed (e.g., Rauch \&
Blandford 1994; Dexter \& Agol 2009; see, however, Broderick 2006;
Dolence et al.\ 2009 for different approaches).

The Kerr metric is very particular both in the sense that it is
completely described by only two parameters (the mass and the spin of
the compact object) and that orbits within this metric are
characterized by the Carter constant (see, e.g., discussion in Will
2009).  Introducing any deviation from the Kerr metric, while
satisfying the vacuum Einstein field equations, does not preserve its
Petrov-type D character and the Carter constant is no longer conserved
along geodesics (see, e.g., Glampedakis \& Babak 2006; Gair, Li, \&
Mandel 2008).  As a result, ray tracing in a non-Kerr metric cannot be
performed entirely using integrals of motions but requires integrating
the second-order differential equations for individual geodesics.

There are at least two distinct astrophysical settings for which ray
tracing in a metric that deviates from the Kerr solution is important.
First, the external spacetime of a neutron star spinning at $\simeq
300-600$~Hz, which is typical of X-ray bursters in low-mass X-ray
binaries, is not accurately described by the Kerr metric. Effects
related to the oblateness of the star (Morsink et al.\ 2007) as well
as to deviations of the quadrupole moment of its spacetime from the
Kerr value (Hartle \& Thorne 1968; see also Laarakkers \& Poisson
1999; Berti \& Stergioulas 2004) are not negligible at these spin
frequencies. Matching the theoretical models to the level of accuracy
reached with current observations of spinning neutron stars can only
be achieved by considering at least the deviation of the quadrupole
moments of their spacetimes from the Kerr values.

Second, calculating the observational appearance of black holes with
arbitrary quadrupole moments can be used in testing the no-hair
theorem with astrophysical observations (Johannsen \& Psaltis 2010a,
2010b, 2010c). The absence of additional `hair' ensures that all
moments of a black hole spacetime that are higher than the dipole have
a particular dependence on the mass and the spin of the black hole. In
particular, the quadrupole moment $q$ of a black hole spacetime has to
depend on its spin $a$ according to the relation $q=-a^2$, where all
quantities have been normalized with appropriate powers of the mass
$M$ of the black hole. Allowing for the quadrupole moment of the
spacetime to take arbitrary values and using observations to test the
validity of the above relation between the quadrupole and the spin of
the black hole constitutes a formal quantitative test of the no-hair
theorem (Ryan 1995).

Here we present a new ray-tracing algorithm for calculating the
observational appearance of spinning compact objects with arbitrary
quadrupole moments. We employ the metric of Glampedakis \& Babak
(2006), which is characterized by three parameters: the mass and spin
of the compact object and its quadrupole moment. In the algorithm, we
integrate two first-order differential equations that arise from
integrals of motion as well as two second-order differential equations
for two components of the geodesic equations in order to compute the
trajectories of photons in these spacetimes. We also use a third
integral of motion related to the norm of the photon 4-momenta in
order to monitor the accuracy of the calculations. In the first paper
of this series, we focus on an application related to the line spectra
of accreting black holes with spacetimes that violate the no-hair
theorem.

In \S2, we describe the metric of Glampedakis \& Babak (2006) in some
detail and, in \S3, we outline the numerical algorithm. Finally, in
\S4, we present some illustrative results for different astrophysical
settings while comparing the results of our algorithm with other
calculations for Kerr black holes.

\section{A Vacuum Metric with an Arbitrary Quadrupole Moment}

We describe the external spacetime of a spinning compact object with
an arbitrary quadrupole moment using the metric of Glampedakis \&
Babak (2006). This metric arises by adding to the Kerr solution a
contribution that has an arbitrary quadrupole moment and is by itself
a solution to the vacuum Einstein field equations (Hartle \& Thorne
1968).

The metric is specified uniquely by three parameters: the mass $M$ of
the compact object, the spin $a$, and the deviation $\epsilon$ of its
quadrupole moment from the Kerr value. Setting $\epsilon=0$ makes the
metric equal to the Kerr solution, which is appropriate for a black
hole of arbitrary spin. On the other hand, when $a/M\lesssim 0.4$, all
moments of the spacetime that are higher than the quadrupole are
negligible and the metric becomes appropriate for neutron stars that
are spinning moderately but not close to their mass shedding limit
(Hartle \& Thorne 1968).

Following Glampedakis \& Babak (2006), we write the metric in
Boyer-Lindquist coordinates as
\begin{equation}
g_{\mu\nu}=g_{\mu\nu}^{\rm K}+\epsilon h_{\mu\nu}\;.
\label{eq:GB}
\label{qKerr}
\end{equation}
Here $g_{\mu\nu}^{\rm K}$ is the Kerr metric with the line element
\begin{equation}
ds^2=-\left(1-\frac{2Mr}{\Sigma}\right)~dt^2
-\left(\frac{4Mar\sin^2\theta}{\Sigma}\right)~dtd\phi
+\left(\frac{\Sigma}{\Delta}\right)~dr^2+\Sigma~d\theta^2
+\left(r^2+a^2+\frac{2Ma^2r\sin^2\theta}{\Sigma}\right)\sin^2\theta~d\phi^2\;.
\label{kerr}
\end{equation}
In this relation,
\begin{equation}
\Delta\equiv r^2-2Mr+a^2,
\end{equation}
and
\begin{equation}
\Sigma\equiv r^2+a^2\cos^2~\theta\;.
\label{deltasigma}
\end{equation}

The quadrupole correction is given, in contravariant form, by
\begin{eqnarray}
h^{tt}&=&(1-2M/r)^{-1}\left[\left(1-3\cos^2\theta\right)
\mathcal{F}_1(r)\right],\nonumber\\
h^{rr}&=&(1-2M/r)\left[\left(1-3\cos^2\theta\right)\mathcal{F}_1(r)\right],
\nonumber\\
h^{\theta\theta}&=&-\frac{1}{r^2}\left[\left(1-3\cos^2\theta\right)
\mathcal{F}_2(r)\right],\nonumber\\
h^{\phi\phi}&=&-\frac{1}{r^2\sin^2\theta}\left[\left(1-3\cos^2\theta\right)
  \mathcal{F}_2(r)\right],\nonumber\\
h^{t\phi}&=&0\;,
\end{eqnarray}
with the functions $\mathcal{F}_{1,2}(r)$ shown explicitly in
Appendix~A of Glampedakis \& Babak (2006). If the spacetime of a
compact object is described by this solution, its quadrupole moment is
\begin{equation}
q=-M(a^2+\epsilon M^2)\;,
\end{equation}
whereas all the higher order moments take their corresponding Kerr
values. 

Calculating the observational appearance of a black hole that violates
the no-hair theorem using the above metric requires a phenomenological
scheme to handle its irregularities at $r\simeq 2M$ (see Glampedakis
\& Babak 2006; Gair et al.\ 2008; Johannsen \& Psaltis
2010a). Precisely because of the no hair theorem, the only
axisymmetric vacuum solution to the Einstein field equations that does
not contain naked singularities or closed time-like loops is the Kerr
metric. Allowing for deviations of the quadrupole moment while
requiring that the spacetime remains a solution to the field equations
is necessarily accompanied by the introduction of pathologies to the
spacetime. In the case of the metric~(\ref{eq:GB}) and for spins
$a/M\lesssim 0.4$, these pathologies appear at $r\le 2.6M$ (Johannsen
\& Psaltis 2010a). We consider these to be unphysical and handle them
by removing from the domain of solution all inbound photons that cross
$r=2.6M$. Because, for $a/M\lesssim 0.4$, this radius is smaller than
the radius of the photon orbit, we expect that this scheme affects
only marginally the images and spectra seen by an observer at
infinity.

\section{The Ray Tracing Algorithm}

In this section, we describe the numerical algorithm for the
calculation of the trajectories of individual photons from the image
plane of an observer at infinity to the location of their
emission. Depending on the problem at hand, the latter may be the
stellar surface or an accretion disk. Following Cadeau et al.\ (2007),
we use two integrals of motion to write first-order differential
equations for the time coordinate and the azimuth of each photon
trajectory. We then complete the system using the second-order
differential equations for the geodesics along the radial and polar
coordinates.

The metric~(\ref{eq:GB}) is stationary and axisymmetric. It is,
therefore, characterized by the two usual Killing vectors,
$\xi=(1,0,0,0)$ and $\eta=(0,0,0,1)$, which correspond to the
conservation of energy
\begin{equation}
E=-g_{tt}\frac{dt}{d\lambda}-g_{t\phi}\frac{d\phi}{d\lambda}
\label{eq:energy}
\end{equation}
and angular momentum
\begin{equation}
L=g_{\phi\phi}\frac{d\phi}{d\lambda}+g_{t\phi}\frac{dt}{d\lambda}
\label{eq:angmomentum}
\end{equation}
along the photon trajectory. Here, $g_{\mu\nu}$ is the
$\mu\nu-$element of the metric, and $\lambda$ is an affine
parameter. Using these two conserved quantities, we now write two
first-order differential equations for the evolution of the $t-$ and
$\phi-$ components of the photon position as
\begin{equation}
\frac{dt}{d\lambda^\prime}=\frac{-g_{\phi\phi}-bg_{t\phi}}
{g_{\phi\phi}g_{tt}-g^2_{t\phi}}
\label{eq:t}
\end{equation}
and
\begin{equation}
\frac{d\phi}{d\lambda^\prime}=\frac{bg_{tt}+bg_{t\phi}}
{g_{\phi\phi}g_{tt}-g^2_{t\phi}}\;.
\end{equation}
where we have defined the normalized affine parameter
$\lambda^\prime\equiv E\lambda$ and the impact parameter for the
photon trajectory $b\equiv L/E$.

For the $r-$ and $\theta-$ components of the photon position we use
the second-order geodesic equations, which for a general axisymmetric 
metric take the form
\begin{equation}
\frac{d^2r}{d\lambda^{\prime 2}}=
-\Gamma^r_{tt}\left(\frac{dt}{d\lambda^\prime}\right)^2
-\Gamma^r_{rr}\left(\frac{dr}{d\lambda^\prime}\right)^2
-\Gamma^r_{\theta \theta}\left(\frac{d\theta}{d\lambda^\prime}\right)^2
-\Gamma^r_{\phi \phi}\left(\frac{d\phi}{d\lambda^\prime}\right)^2
-2\Gamma^r_{\phi t}\left(\frac{d\phi}{d\lambda^\prime}\right)
\left(\frac{dt}{d\lambda^\prime}\right)
-2\Gamma^r_{\theta r}\left(\frac{d\theta}{d\lambda^\prime}\right)
\left(\frac{dr}{d\lambda^\prime}\right)
\label{eq:r}
\end{equation}
and
\begin{equation}
\frac{d^2\theta}{d\lambda^{\prime 2}}=
-\Gamma^\theta_{tt}\left(\frac{dt}{d\lambda^\prime}\right)^2
-\Gamma^\theta_{rr}\left(\frac{dr}{d\lambda^\prime}\right)^2
-\Gamma^\theta_{\theta \theta}\left(\frac{d\theta}{d\lambda^\prime}\right)^2
-\Gamma^\theta_{\phi \phi}\left(\frac{d\phi}{d\lambda^\prime}\right)^2
-2\Gamma^\theta_{\phi t}\left(\frac{d\phi}{d\lambda^\prime}\right)
\left(\frac{dt}{d\lambda^\prime}\right)
-2\Gamma^\theta_{\theta r}\left(\frac{d\theta}{d\lambda^\prime}\right)
\left(\frac{dr}{d\lambda^\prime}\right)
\label{eq:theta}
\end{equation}
Here, $\Gamma^\alpha_{\beta\gamma}$ are the various Christoffel
symbols for the metric~(\ref{eq:GB}).

A final integral of motion arises from the requirement that the norm of the
photon 4-momentum has to vanish, i.e.,
\begin{equation}
g_{tt}\left(\frac{dt}{d\lambda^\prime}\right)^2
+g_{rr}\left(\frac{dr}{d\lambda^\prime}\right)^2
+g_{\theta\theta}\left(\frac{d\theta}{d\lambda^\prime}\right)^2
+g_{\phi\phi}\left(\frac{d\phi}{d\lambda^\prime}\right)^2
+2g_{t\phi}\left(\frac{dt}{d\lambda^\prime}\right)
\left(\frac{d\phi}{d\lambda^\prime}\right)=0\;.
\label{eq:4mom}
\end{equation}
This integral of motion is not useful for replacing either the geodesic
equation~(\ref{eq:r}) or (\ref{eq:theta}) because it contains the
squares of the derivatives of the $r-$ and $\theta-$ coordinates with
respect to the affine parameter. Keeping track of the appropriate sign
for the two derivatives, especially near the inflection points of the
geodesics, would more than offset the benefit of using a first-order
integral of motion as opposed to a second-order geodesic
equation. Therefore, following Cadeau et al.\ (2007), we use this
integral of motion only in order to monitor the accuracy of the
calculation. To this end, we define the parameter
\begin{equation}
\xi\equiv 
\left[g_{rr}\left(\frac{dr}{d\lambda^\prime}\right)^2
+g_{\phi\phi}\left(\frac{d\phi}{d\lambda^\prime}\right)^2
+g_{\theta\theta}\left(\frac{d\theta}{d\lambda^\prime}\right)^2
+2g_{t\phi}\left(\frac{dt}{d\lambda^\prime}\right)
\left(\frac{d\phi}{d\lambda^\prime}\right)\right]/
\left[g_{tt}\left(\frac{dt}{d\lambda^\prime}\right)^2\right]
\end{equation}
and test whether its value remains equal to $\xi=-1$ along each
geodesic.

Starting from a fine raster of points on the image plane of an
observer at infinity, we follow the geodesics backwards to the surface
of the compact object or to different regions in the accretion flow
where the photons originate. 

\begin{figure}[t]
\centerline{\psfig{file=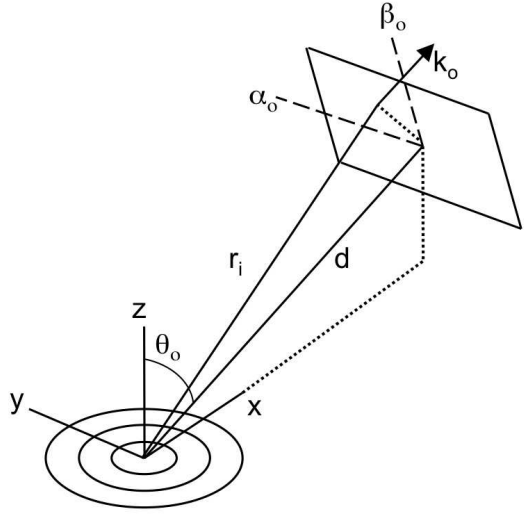,width=3.0in}}
\caption{The geometry of ray tracing.}
\label{fig:geometry}
\end{figure}

Following Johannsen \& Psaltis (2010b), we consider an observer
viewing the central object from a large distance $d$ and at an
inclination angle $\theta_o$ from its rotation axis (see
Fig.~\ref{fig:geometry}). We set up a virtual image plane that is
perpendicular to the line of sight and centered at $\phi=0$ of the
spacetime.

We define the set of Cartesian coordinates $(\alpha_0,\beta_0)$ on the
image plane such that the $\beta_0$-axis is along the same fiducial
plane and the $\alpha_0$-axis is perpendicular to it. We then convert
the coordinates $(\alpha_0,\beta_0)$ of a photon that reaches the
image plane to the coordinates $(r_i,\theta_i, \phi_i)$ in the
spherical-polar system used for the metric~(\ref{eq:GB}) with the
relations (see Johannsen \& Psaltis 2010b)
\begin{eqnarray}
r_i&=&\left(d^2+\alpha_0^2+\beta_0^2\right)^{1/2}\\
\cos\theta_i&=&\frac{1}{r_i}\left(d\cos\theta_o+\beta_0\sin\theta_o\right)\\
\tan\phi_i&=&\alpha_0\left(d\sin\theta_o-\beta_0\cos\theta_o\right)^{-1}\;.
\end{eqnarray}

The photons that contribute to the image of the compact object are those
with 3-momenta that are perpendicular to the image plane. This orthogonality
condition uniquely specifies the momentum vector of a photon with the 
above coordinates, according to the relations (Johannsen \& Psaltis 2010b)
\begin{eqnarray}
k^r&\equiv&\frac{dr}{d\lambda^\prime}=\frac{d}{r_i}\\
k^\theta&\equiv&\frac{d\theta}{d\lambda^\prime}=
\left[-\cos\theta_o+\frac{d}{r_i^2}\left(d\cos\theta_o+\beta_0\sin\theta_o\right)
\right]
\left[r_i^2-(d\cos\theta_o+\beta_0\sin\theta_o)^2\right]^{-1/2}
\label{eq:ktheta}\\
k^\phi&\equiv&\frac{d\phi}{d\lambda^\prime}=
\frac{-\alpha_0\sin\theta_o}{(d\sin\theta_o-\beta_0\cos\theta_o)^2+\alpha_0^2}\;.
\label{eq:kphi}
\end{eqnarray}
Using these relations, we then calculate the $t-$component of the
photon 4-momentum from equation~(\ref{eq:4mom}). At this point, the
normalization of the photon 4-momentum is arbitrary. Note that, for a
distant observer, $d/r_i\rightarrow 1$, which implies that special care
needs to be taken in evaluating expressions~(\ref{eq:ktheta}) and 
(\ref{eq:kphi}) to avoid round-off errors.

\begin{figure}[t]
\centerline{\psfig{file=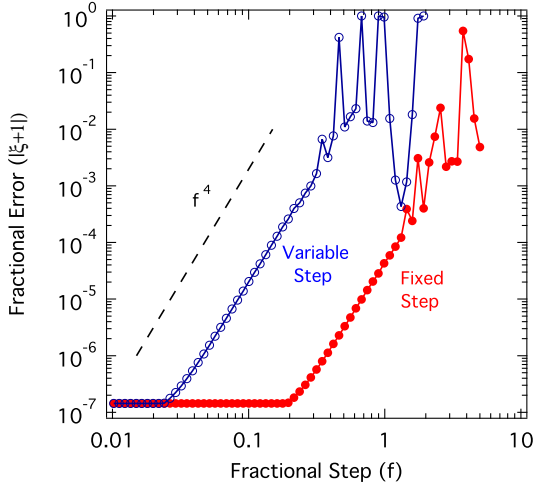,width=3.0in}}
\caption{The fractional error introduced by the integration algorithm
  during the tracing of a single ray from the image plane to the
  equatorial plane, as a function of the fractional stepsize of the
  integration. The image plane is set at a distance of $1028M$ and at
  an inclination of $40^\circ$. The black hole spin is equal to
  $a/M=0.5$ and its spacetime is described by the Kerr metric. The
  Cartesian coordinates of the position of the ray on the image plane are
  $(\alpha_0,\beta_0)=(5,0)$. The results of two integrations are shown,
  one with a fixed step in the affine parameter and one with a
  variable step.  In both cases, the performance of the algorithm is
  consistent with the fourth-order of the integration method.}
\label{fig:convergence}
\end{figure}

In the numerical algorithm, we integrate
equations~(\ref{eq:t})-(\ref{eq:theta}) using a fourth-order
Runge-Kutta integrator. Figure~\ref{fig:convergence} shows the
convergence of our algorithm, for two cases. In the first case, we
employ a fixed step $f$ in the affine parameter. In the second case,
we employ an adaptive stepsize that we set equal to a fixed fraction
$f$ of the inverse rate of the fastest changing variable at each
point, i.e.,
\begin{equation}
\delta \lambda^\prime=f \min\left[
t\left(\frac{dt}{d\lambda^\prime}\right)^{-1},
r\left(\frac{dr}{d\lambda^\prime}\right)^{-1},
\theta\left(\frac{d\theta}{d\lambda^\prime}\right)^{-1},
\phi\left(\frac{d\phi}{d\lambda^\prime}\right)^{-1}
\right]\;.
\end{equation}
In both cases, the algorithm shows the expected fourth-order
convergence of our integration method. For very small values of the
step size, the fractional error stabilizes and reflects the numerical
accuracy to which the parameter $\xi$ was calculated on the image
plane. For the calculations shown hereafter, we will use an adaptive
step with $f=1/32\simeq 0.03$.

Our numerical algorithm is capable of integrating $\simeq 10^4$
geodesics in a time comparable to a second, on a personal computer
with a 2.5~GHz Intel core. This is comparable to the speed of other
similar algorithms that employ different methods (e.g., Dexter \& Agol
2009; Doelence et al.\ 2009). Moreover, the algorithm is trivially
parallelizable and, because of its very low storage requirements, is 
optimal for implementation on a GPU. 

\section{Relativistically Broadened Fluorescent Lines 
Around Quasi-Kerr Black Holes}

As a first application of our numerical algorithm, we calculate the
profiles of relativistic broadened fluorescent iron lines from
geometrically thin accretion disks around quasi-Kerr black holes.
These are prime targets for current and future X-ray telescopes and
are expected to lead to the measurements of the spins of a large
number of black holes in binary systems and in active galactic nuclei
(for reviews see Reynolds \& Nowak 2003; Miller 2007).

For this application, we trace rays from the image plane to the
equatorial plane, where we will assume that a geometrically thin disks
exists, from some outer radius $r_{\rm out}$ down to the radius of the
innermost stable circular orbit (ISCO).

The character of the ISCO in the metric~(\ref{eq:GB}) depends on the
sign and magnitude of the quadrupole deviation parameter $\epsilon$.
When $\epsilon>0$, orbits close to the central object become unstable
to radial perturbations. Following Johannsen \& Psaltis (2010c), we
calculate the location of the ISCO in this case by finding the zero in
the radial profile of the square of the radial epicyclic frequency for
a particle in a circular equatorial orbit.  For values of the
quadrupole deviation parameter $\epsilon$ that are sufficiently
negative, all circular equatorial orbits become stable to radial
perturbations. However even in this case, orbits very close to the
central object become unstable to vertical perturbations (see also
Gair et al.\ 2008).  Presently, we consider only positive deviations
of the quadrupole moment of the metric from the Kerr value, i.e.,
$\epsilon>0$. When $\epsilon=0$, we use the complete expression for
the location of the ISCO from Bardeen et al.\ (1973).

We assume that the disk is composed of a set of equatorial concentric
rings, in which the plasma is moving at the local Keplerian velocity
with (Glampedakis \& Babak 2006)
\begin{eqnarray}
u^\phi&=&\frac{1}{\Delta}
\left[\frac{2M}{r}\left(aE-L\right)+L\right]
-\epsilon \frac{h_3}{r^2}L\\
u^t&=&\frac{1}{\Delta}
\left[E\left(r^2+a^2\right)+\frac{2Ma}{r}(aE-L)\right]
-\epsilon f_3E\left(1-\frac{2M}{r}\right)^{-1}\;.
\end{eqnarray}
Here, $E$ and $L$ are the energy and angular momentum of the circular
orbit at radius $r$ and are given in Johannsen \& Psaltis (2010a).  

\begin{figure}[t]
\centerline{\psfig{file=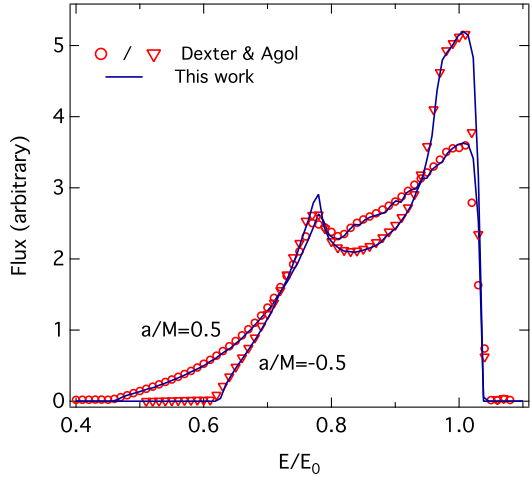,width=3.0in}}
\caption{Profiles of relativistically broadened fluorescent lines
  from a geometrically thin accretion disk around a Kerr black hole,
  for two values of the black-hole spin. The outer radius of the disk
  is set to $15 M$, the power-law index of the line emissivity to
  $\alpha=2$, and the inclination to the observer to $\theta_{\rm
    o}=30^\circ$. The symbols are the results reported by Dexter \&
  Agol (2009), while the solid lines are the profiles calculated with
  the algorithm described here.}
\label{fig:lines_spin}
\end{figure}

\begin{figure}[t]
\centerline{\psfig{file=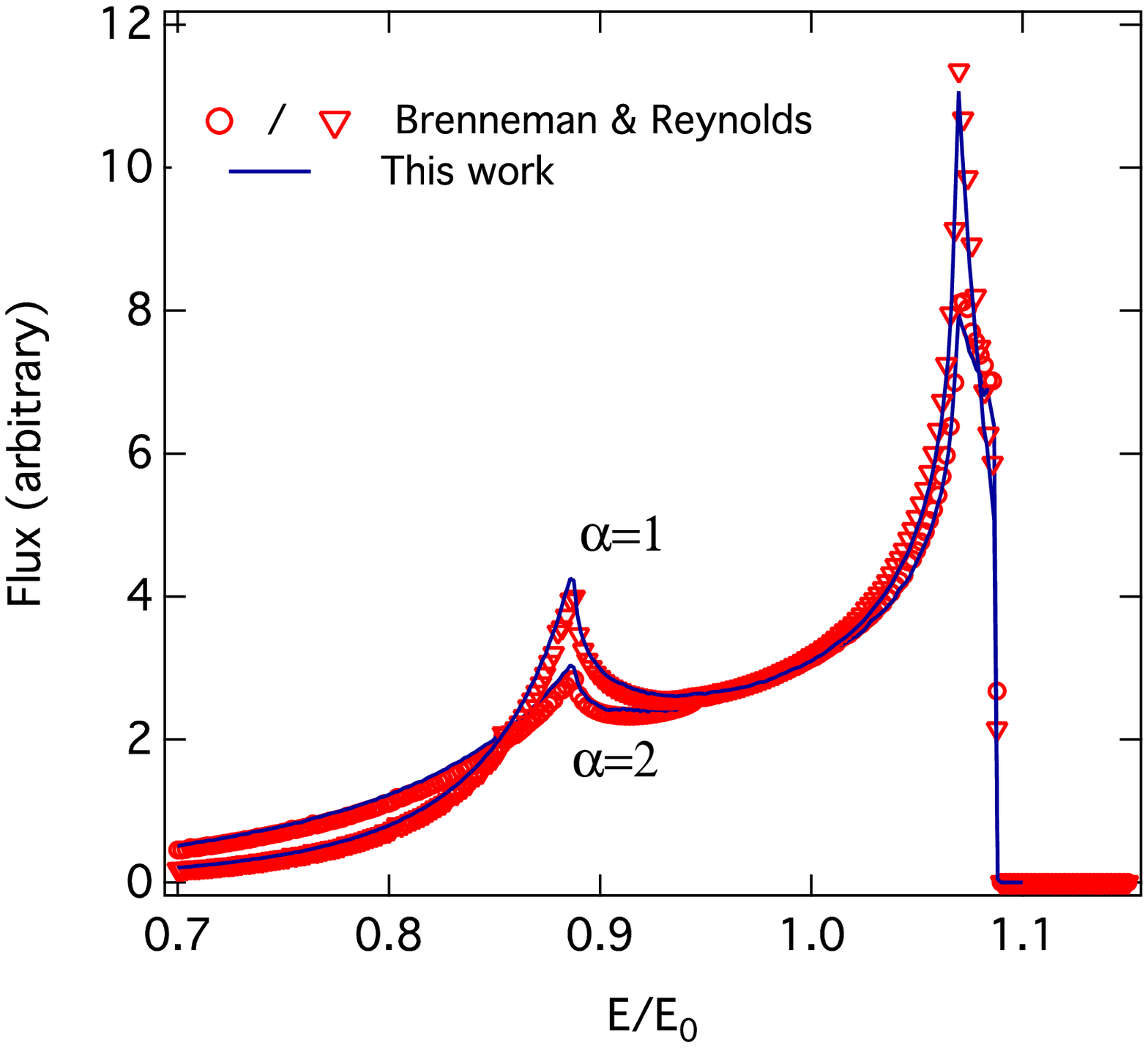,width=3.0in}}
\caption{Profiles of relativistically broadened fluorescent lines from
  a geometrically thin accretion disk around a Kerr black hole, for
  two values of the line emissivity. The black hole spin is equal to
  $a/M=0.5$, the outer radius of the disk is set to $50 M$, and the
  inclination to the observer to $\theta_{\rm o}=40^\circ$. The
  symbols are the results reported by Brenneman \& Reynolds (2006),
  while the solid lines are the profiles calculated with the algorithm
  described here.}
\label{fig:lines_emiss}
\end{figure}

We also assume that radiation emerges from the accretion disk surface
isotropically with an ``emissivity'' that scales as the power-law
function $r^{-\alpha}$ of the coordinate radius. 
We calculate the overall change in the energy of the photon
from the location of emission at the accretion disk to the image plane
using
\begin{equation}
\frac{E_{\rm im}}{E_{\rm d}}=\frac{g_{\mu\nu,{\rm im}}k_{\rm im}^\mu u_{\rm im}^\nu}
{g_{\mu\nu,{\rm d}}k_{\rm d}^\mu u_{\rm d}^\nu}
\end{equation}
and setting the 3-velocity of the observer at the image plane to zero.
In this last expression, the subscripts ``im'' and ``d'' refer to the
image plane and the accretion disk, respectively. We finally employ
the Lorentz invariant quantity $I/E^3$, where $I$ is the monochromatic
specific intensity of the radiation field and $E$ is the photon
energy, in order to calculate the specific intensity at each point on
the image plane.

The result at the completion of this calculation is the overall
redshift or blueshift experienced by a photon that reaches each point
on the image plane, which we denote by $g(\alpha_0,\beta_0)\equiv E_{\rm
  im}/E_{\rm d}$, and the corresponding specific intensity
$I(\alpha_0,\beta_0)$. The monochromatic flux at the image plane is then
\begin{equation}
F_E\sim\frac{1}{d^2}\int d\alpha_0\int d\beta_0 I(\alpha_0,\beta_0)
\delta\left[E-E_0 g(\alpha_0,\beta_0)\right]\;.
\label{eq:flux}
\end{equation}
The presence of the $\delta$-function allows us in principle to
convert this 2-dimensional integral into a one-dimensional line
integral along contours of constant values of the quantity
$g(\alpha_0,\beta_0)$ on the image plane. In practice, calculating the
location of these contours is time consuming. Instead, we evaluate
expression~(\ref{eq:flux}) using a Monte Carlo integration of points
on the image plane and a set of fine bins in photon energy.

\begin{figure}[t]
\centerline{\psfig{file=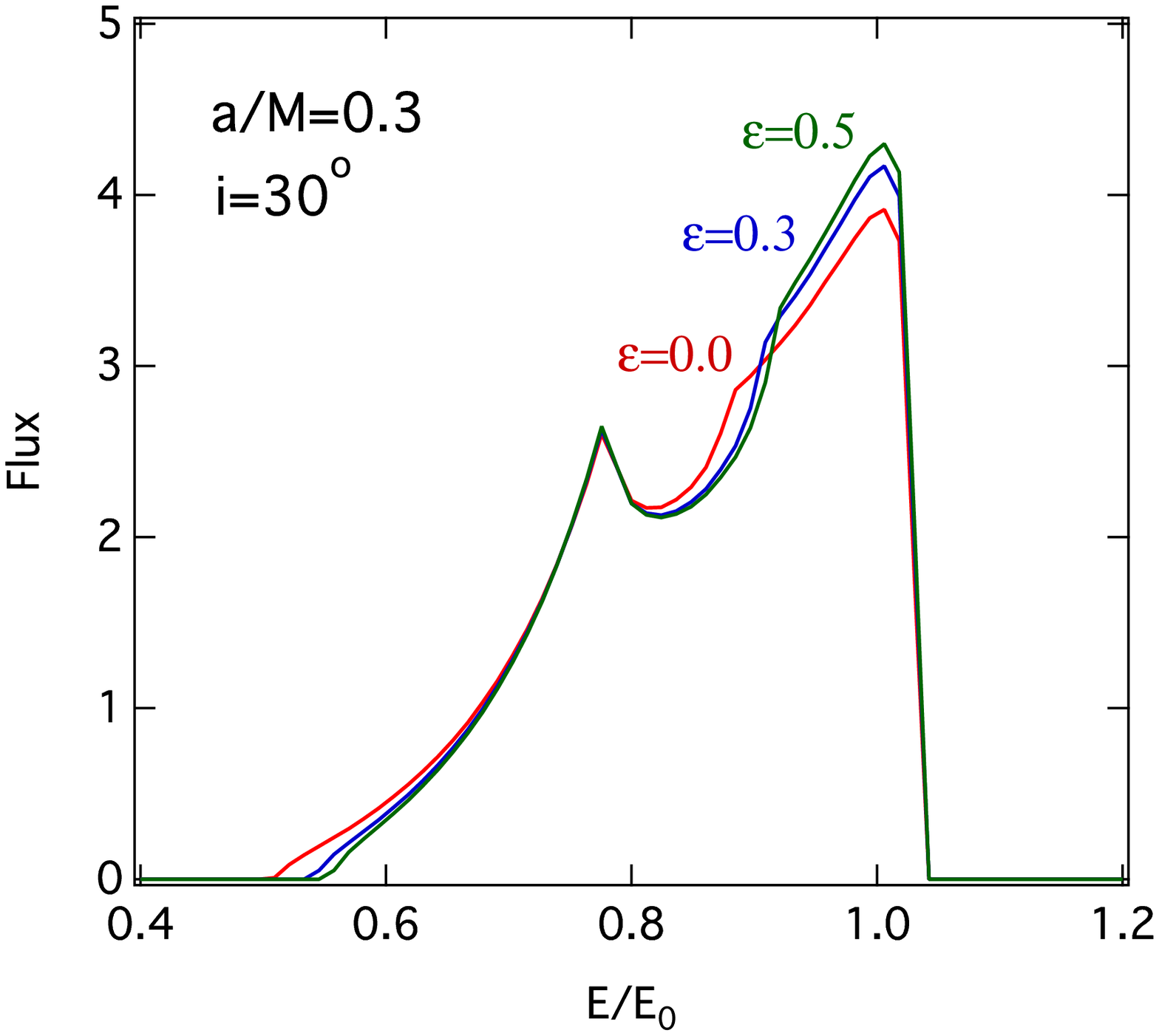,width=3.0in}}
\caption{Profiles of relativistically broadened fluorescent lines
  from a geometrically thin accretion disk around a quasi-Kerr black
  hole, for a spin of $a/M=0.3$ and three values of the quadrupole
  deviation parameter $\epsilon$. The remaining parameters are the
  same as in Figure~\ref{fig:lines_spin}.}
\label{fig:lines_quad}
\end{figure}

Figures~\ref{fig:lines_spin} and \ref{fig:lines_emiss} show the
dependence of the line profiles on the spin of the black hole and on
the emissivity of fluorescence on the accretion disk. They also
compare our results with other calculations from the literature,
demonstrating the agreement between the different numerical
algorithms.  Figure~\ref{fig:lines_quad} shows the dependence of the
fluorescent line profiles on the parameter $\epsilon$, which measures
the deviation of the quadrupole moment of the black-hole spacetime
from its Kerr value. As we increase the degree of quadrupole
deviation, the red wing of the line becomes less pronounced and the
relative strengths of the blue and red wings is altered. Both effects
are predominantly caused by the increase in the radius of the ISCO
with increasing value of the parameter $\epsilon$.

\section{Conclusions}

We described a new ray tracing algorithm for the calculation of 
observables from compact objects with spacetimes characterized by 
arbitrary quadrupole moments. Such spacetimes are relevant to black 
holes that violate the no-hair theorem and to moderately spinning 
neutron stars. We put special care in streamlining and accelerating
our algorithm in order to achieve the efficiency neccessary for large
parameter studies and comparisons to data. We also demonstrated the
expected convergence of our algorithm and verified our results against
those of previous calculations for Kerr metrics.

As a first application, we calculated the profiles of fluorescent iron
lines from black holes that violate the no-hair theorem. As expected,
varying the quadrupole moment of the spacetime, led to changes in the
detailed profiles of the lines.  We will study the observability of
these effects as well as ways of breaking the degeneracy between
changing the spin and the quadrupole of the black hole in a
forthcoming paper.

\bibliographystyle{apj}

\end{document}